\documentclass[aps,prl,twocolumn,superscriptaddress]{revtex4-2}

\usepackage{amsmath}
\usepackage{graphicx}
\usepackage{siunitx}
\usepackage[T1]{fontenc}
\usepackage{layouts}

\begin{document}
\title{Impact of transforming interface geometry on edge states in valley photonic crystals}
\author{D. Yu}
\thanks{These authors contributed equally.}
\affiliation{Kavli Institute of Nanoscience, Delft University of Technology, 2600 GA, Delft, The Netherlands}
\author{S. Arora}
\thanks{These authors contributed equally.}
\affiliation{Kavli Institute of Nanoscience, Delft University of Technology, 2600 GA, Delft, The Netherlands}
\author{L. Kuipers}
\email{l.kuipers@tudelft.nl}
\affiliation{Kavli Institute of Nanoscience, Delft University of Technology, 2600 GA, Delft, The Netherlands}

\date{\today}

\begin{abstract}
Topologically protected edge states arise at the interface of two topologically distinct valley photonic crystals. In this work, we investigate how tailoring the interface geometry, specifically from a zigzag interface to a glide plane, profoundly affects these edge states. Near-field measurements demonstrate how this transformation significantly changes the dispersion relation of the edge mode. We observe a transition from gapless edge states to gapped ones, accompanied by the occurrence of slow light within the Brillouin zone, rather than at its edge. Additionally, we simulate the propagation of the modified edge states through a specially designed valley-conserving defect. The simulations show, by monitoring the transmittance of this defect, how the robustness to backscattering gradually decreases, suggesting a disruption of valley-dependent transport. These findings demonstrate how the gradual emergence of valley-dependent gapless edge states in a valley photonic crystal depends on the geometry of its interface.
\end{abstract}

\maketitle


The discovery of topological phases in photonic crystals provides a new degree of freedom to manipulate light-matter interaction. Photonic crystals with nontrivial topological phases are called topological photonic crystals (TPCs) \cite{Raghu2008a}. TPCs support nontrivial edge states that interface different topological phases. These edge states are known as topological edge states (TESs). TESs feature gapless dispersion and robust transport characteristics \cite{ma2015guiding,khanikaev2017two}, making them promising for broadband lossless on-chip communication \cite{guglielmon2019broadband,ma2019topological,yang2020terahertz,tan2022interfacial}. The bulk-boundary correspondence relates the TESs to the topological order in the bulk of TPCs \cite{mong2011edge,Ozawa2019a}. By engineering the bulk of TPCs, TESs associated with various topological orders have been realized with various degrees of robustness \cite{Raghu2008a,rechtsman2013photonic,Wu2015,ma2016all,xi2020topological,davis2022topologically}. Additionally, the interface geometry also plays an important role in engineering the edge states. Proper design of the interface can lead to, e.g., chiral interface \cite{mehrabad2020chiral} or broadband low-loss waveguides \cite{tan2022interfacial,wen2022designing}.

The effect of transforming the interface geometry on TESs depends on the type of TPCs. For nonreciprocal TPCs, such as gyromagnetic photonic crystals \cite{Wang2008,Wang2009}, the existence of TESs is guaranteed by the topological order in bulk (i.e., topological protection) \cite{Wang2008,haldane2008possible}. Therefore, the precise interface structure has little impact on the TESs \cite{Wang2008}. For $C_6$-symmetric TPCs \cite{Wu2015,yang2018visualization,parappurath2020direct}, the topological protection of TESs is conditional on the conservation of pseudo-spin \cite{Wu2015}. If an interface mixes different pseudo-spins, it will be detrimental to the gapless dispersion and the robustness of TESs \cite{Wu2015}. A similar argument applies to valley photonic crystals (VPCs), where the valley degree of freedom plays the role of pseudo-spin \cite{ma2016all,Dong2017}. Certain interfaces of VPCs can couple different valleys and are unable to support TES \cite{noh2018observation,Wong2020}. Conversely, it is widely accepted that an interface of two topologically distinct VPCs should exhibit TESs as long as it respects the conservation of valleys \cite{ma2016all,Shalaev2019a,xue2021topological}. This statement implies that edge states of VPCs are robust against perturbations on the interface geometry. However, recent studies have demonstrated that modifying the interface of VPCs can significantly influence the properties of edge states \cite{chen2018tunable, yoshimi2020slow,devi2021topological,li2022valley}. This finding urges us to reassess the robustness of these edge states to perturbations on the interface.

Here, we study the impact of gradually changing the geometry of a VPC interface on edge states in a valley photonic crystal, with a specific focus on the transformation from a zigzag interface into a glide plane. We fabricate VPCs with these tailored interfaces and exploit near-field optical microscopy to map the wavefunctions of their edge states. The measurements demonstrate that the edge states undergo a transition from gapless to gapped, even when the conservation of valleys is preserved by the transformation. Meanwhile, we observe that edge states slow down within the Brillouin zone (BZ), in contrast with typical valley-dependent edge states that only become slow at the edge ($k_x=\pm\pi/a$) or in the center ($k_x=0$) of the BZ \cite{Shalaev2019a,Arora2021}. Next, we examine the valley-dependent transport of these edge states by simulating their propagation through a specially designed valley-conserving defect. The calculated transmittance of this defect experiences a significant drop as the geometry of the interface is transformed. This observation suggests that the transformation of the VPC interface disrupts the valley-dependent transport of the edge states.

\begin{figure*}[htbp]
    
    \includegraphics{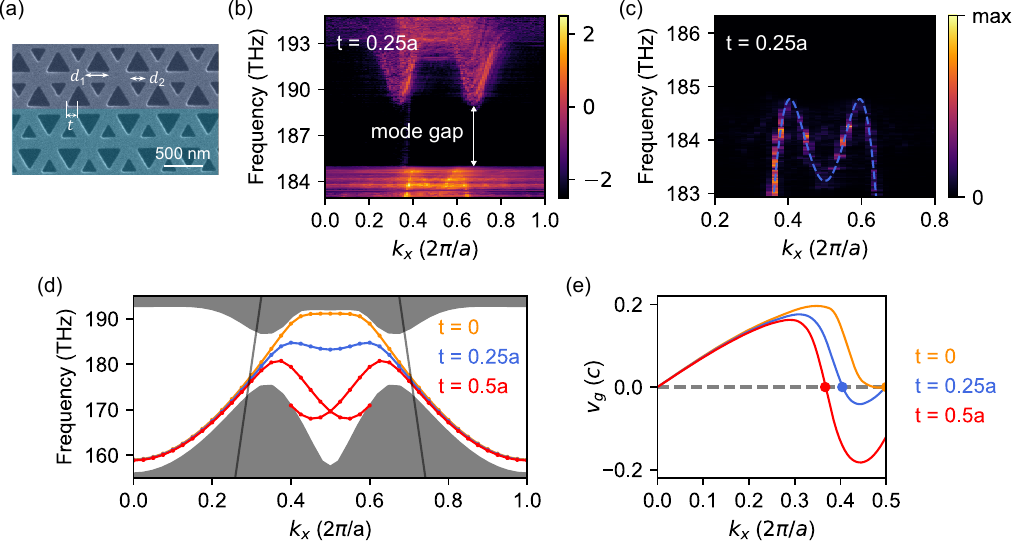}
        \caption{\label{fig:dispersion} Interface geometry and dispersion diagrams. (a) An SEM image of a shifted interface of two distinct VPCs for $t=0.25a$, with $a=\SI{500}{\nano\meter}$ the lattice constant of both VPCs. (b) Experimentally retrieved dispersion diagram of photonic modes of the shifted interface, where $\log_{10}|\mathcal{F}(\mathbf{E})|^2$ is plotted versus $k_x$ and the frequency of excitation. A mode gap arises between the edge states (bottom of graph) and the bulk modes (top of graph). A close-up of the measured dispersion curve of the edge states for a shift of $0.25a$ is shown in (c), where $|\mathcal{F}(\mathbf{E})|^2$ is plotted in a linear scale. The measured dispersion has an M-shape and is consistent with the simulation result (blue-dashed line). Remarkably, there is a slow light region around $k_x=0.8\pi/a$, which is inside the Brillouin zone rather than at its edge. As a result, the edge state also exhibits both a positive and negative group velocity within half a Brillouin zone. (d) Numerically simulated dispersion curves of edge states for $t=0$ (orange), $0.25a$ (blue), and $0.5a$ (red). The grey regions represent bulk modes, and the dark line corresponds to the light line $\omega=ck$. As $t$ increases from $0$ to $0.5a$, the edge states transform from gapless to gapped. (e) The group velocity of edge states. With an increase in $t$, a slow-light edge state with $v_g=0$ transitions from the Brillouin zone edge towards its interior.}
\end{figure*}

We start with a VPC interface, as depicted in Fig.~\ref{fig:dispersion}(a), where two distinct VPCs are patterned on a silicon-on-insulator slab. Each unit cell of these VPCs comprises two inequivalent triangular holes, with side lengths of $d_1=\SI{300}{\nano\meter}$ and $d_2=\SI{200}{\nano\meter}$, respectively. The lattice constant of both VPCs is $a=\SI{500}{\nano\meter}$. The two VPCs are inversion images of each other, resulting in a difference in valley-dependent topology between them \cite{Shalaev2019a}. To modify the interface geometry, we shift one VPC along the direction of the interface by a distance of $t$ (glide). When $t=0$, we call this interface a zigzag interface, which is mirror-symmetric and supports valley-dependent gapless edge states \cite{ma2016all,Shalaev2019a,Arora2021}. As $t$ increases to $0.5a$, the interface geometry becomes glide-plane symmetric and has been transformed into a glide plane. In particular, we refer to the interface with $t=0.25a$ as the shifted interface, which represents the middle point of the transformation. It is important to note that shifting the VPC preserves the bulk symmetry and, as a result, does not affect the valley-dependent topology. Moreover, this perturbation also respects the conservation of valleys since it does not influence the wavefunction overlap between states at different valleys \cite{ma2016all}. Consequently, we expect valley-dependent gapless edge states to appear both at the shifted interface and the glide plane.

We fabricate three VPC interfaces, corresponding to $t=0$, $0.25a$, and $0.5a$, respectively. The in-plane electric field distribution $\mathbf{E}$ over these interfaces is measured with phase-sensitive near-field scanning optical microscopy (NSOM) \cite{Rotenberg2014}. We apply a spatial Fourier transform to the measured complex electric field $\mathbf{E}$, denoted by $\mathcal{F}(k_x)$, and repeat this process for all wavelengths to obtain the dispersion relation of the photonic modes \cite{Arora2021}. Due to the periodic nature of $\mathcal{F}(k_x)$ in reciprocal space, the data from a single BZ is adequate for retrieving all dispersion curves. Nevertheless, to enhance the signal-to-noise ratio, we employ BZ folding by summing the intensities of all Bloch harmonics, which involves data from all BZs. This technique yields the final dispersion diagrams of photonic modes of the VPC interfaces. For $t=0$, we observe edge states with a gapless dispersion curve for the zigzag interface (see Fig.~S1 in Supplemental Material \cite{supplemental}), as expected from previous studies \cite{Shalaev2019a,Arora2021}. However, as $t$ is increased the dispersion curve changes dramatically. The retrieved dispersion diagram of photonic modes of the shifted interface is shown for $t=0.25a$ in Fig.~\ref{fig:dispersion}(b), where $|\mathcal{F}(\mathbf{E})|^2$ is plotted versus $k_x$ and excitation frequency (in order to present all relevant features a logarithmic color scale is used). Significantly, we observe that a mode gap has opened up between \SI{185}{\tera\hertz} and \SI{189}{\tera\hertz}, which is absent in the zigzag interface. Thus, the interface deformation by a longitudinal shift can transform gapless edge states into gapped ones. A close-up of the dispersion curve for the edge states is displayed in Fig.~\ref{fig:dispersion}(c), where $|\mathcal{F}(k_x)|^2$ is shown with a linear scale. The measured dispersion curve has an M-shape centering at the BZ edge, which is consistent with the simulation result indicated by the dashed line. This dispersion curve has a slope of zero at approximately $k_x=0.8\pi/a$, demonstrating that slow light occurs within the BZ, namely $k_x<\pi/a$. This slow light region is distinct from the typical ones in a photonic crystal waveguide, which usually lie either at the edge ($k_x=\pi/a$) or in the center ($k_x=0$) of the BZ \cite{krauss2007slow,Yoshimi2020}. Such nontrivial slow light arises from the emergence of energy vortexes, leading to a decrease in group velocity. For further information, refer to \cite{sukhorukov2009slow}. In addition, we present the measured dispersion diagram of photonic modes of the glide plane for $t=0.5a$ (Fig.~S1 \cite{supplemental}), where edge states are not observed as they fall below the operating frequency range of our laser.

To investigate how the dispersion relation of edge states changes with the glide of VPC interfaces, we numerically simulate the dispersion relations of edge states using COMSOL Multiphysics® software \cite{multiphysics1998introduction}, as shown in Fig.~\ref{fig:dispersion}(d). We also present the calculated group velocities of the edge states in Fig.~\ref{fig:dispersion}(e). At the zigzag interface ($t=0$), the edge states exhibit a gapless dispersion curve, which becomes almost linear near the valley at $k_x=2\pi/3a$. After a glide of $t=0.25a$ is applied, a mode gap appears between the edge states and the upper bulk modes. The group velocity of the edge states reduces around the BZ edge ($k_x=\pi/a$), creating a slow light region within the BZ. As $t$ increases to $0.5a$, the mode gap widens further, and the slow light region gets closer to the valley. More calculated dispersion curves are given in Fig.~S2 \cite{supplemental}, underpinning the gradual change from gapless to gapped edge states. In summary, transforming the zigzag interface with a glide causes a transition from gapless edge states to gapped ones, with slow light occurring within the BZ.

\begin{figure*}[htbp]
    \includegraphics{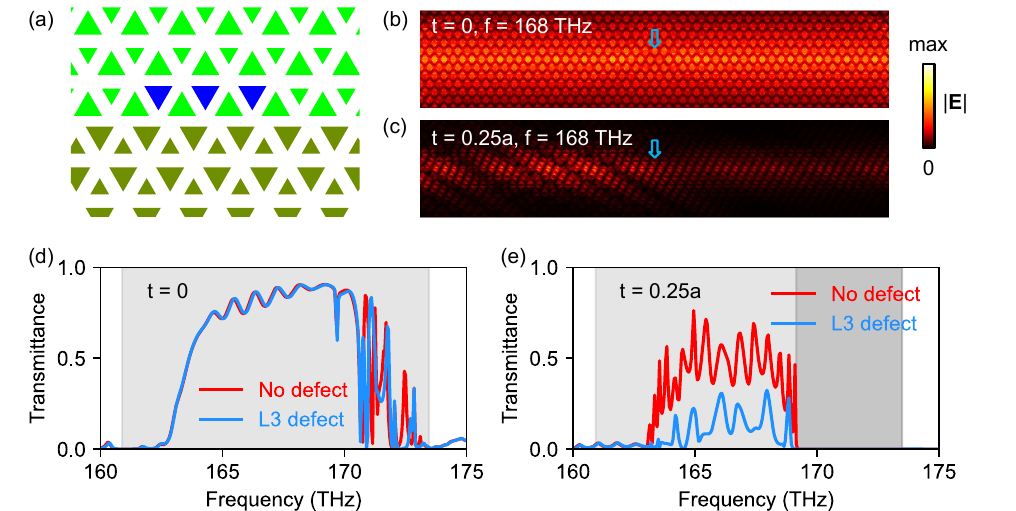}
    \caption{\label{fig:transmission} Defect and transmittance. (a) The geometry of a valley-conserving L3 defect introduced into a shifted interface, with three small triangles replaced by three large ones (blue). (b) \& (c) Simulated electric field amplitudes $|\mathbf{E}|$ on the zigzag interface ($t=0$) and the shifted interface ($t=0.25a$), respectively, at an excitation frequency of \SI{168}{\tera\hertz}. The blue arrow indicates the position of the L3 defect. Light enters the VPC interface from the left side and exits from the right side. A decrease in $|\mathbf{E}|$ at the L3 defect and an interference pattern are observed in (c) but not in (b), indicating a significant contrast of transmittance of that defect between the two VPC interfaces. The simulated transmittance spectra of the zigzag interface and the shifted interface are presented in (d) \& (e), respectively. The grey domain indicates the band gap, while the dark one represents the mode gap. The zigzag interface has a transmittance spectrum almost unaffected by the L3 defect. In contrast, the shifted interface exhibits a significant decreases in its transmittance after introducing the L3 defect.}
\end{figure*}

We found that the edge states at the shifted interface have a dispersion relation distinct from those of typical valley-dependent edge states, which cross the band gap around valleys. Usually, valley-dependent transport is demonstrated with backscattering-free propagation through a valley-conserving defect, such as a sharp waveguide bend at a 120-degree angle \cite{Shalaev2019a,He2019,Arora2021}. However, in the case of the shifted interface, a sharp waveguide bend cannot be realized while preserving the interface geometry. To examine the valley-dependent transport of edge states at this interface, we propose a valley-conserving lattice defect that we call an L3 defect. This defect is introduced into a VPC interface by replacing three small triangular holes with three larger ones, as depicted in Fig.~\ref{fig:transmission}(a). According to first-order perturbation theory, the L3 defect conserves the valley degree of freedom and is thus ideally suited for testing valley-dependent transport (a mathematical proof is provided in Supplemental Material \cite{supplemental}).

We simulate light propagation along our various VPC interfaces using the software mentioned earlier, both with and without an embedded L3 defect. Fig.~\ref{fig:transmission}(b) and \ref{fig:transmission}(c) show the amplitude of the simulated normalized electric field $|\mathbf{E}|$ on the zigzag and shifted interfaces, respectively, for an excitation frequency of $f=\SI{168}{\tera\hertz}$. The input port and output ports are located on the left and right sides, respectively. The arrow indicates the position of the L3 defect. For the zigzag interface, the electric field amplitude remains almost unchanged after passing the L3 defect, indicating the absence of backscattering at that defect. However, for the shifted interface, the electric field amplitude reduces significantly after light passes the L3 defect, indicating the occurrence of scattering at the defect. The interference pattern in Fig.~\ref{fig:transmission}(c), which is not seen in Fig.~\ref{fig:transmission}(b), before the defect demonstrates that most of the scattering is actually backscattering.

The transmittance spectra of the VPC interfaces for two cases, namely $t=0$ (zigzag interface) and $t=0.25a$ (shifted interface), are presented in Fig.\ref{fig:transmission}(d) and (e), respectively. In these figures, the blue line represents the transmittance of the VPC interfaces with an L3 lattice defect, while the red line corresponds to the transmittance without the L3 lattice defect. The grey domain represents the band gap, whereas the dark domain represents the mode gap. Within the band gap and outside the mode gap, the transmittance values are considerably high, indicating the propagation of light through edge states of the VPC interfaces. Furthermore, all transmittance curves exhibit oscillating behavior at various levels. These oscillations might stem from the coupling loss at the input and output ports (not shown in the figure), which makes the transmittance without defects frequency-dependent. It is worth mentioning that the band gap, which spans from \SI{161}{THz} to \SI{173}{THz} (see Fig.~S3 \cite{supplemental}), does not match the one displayed in Fig.\ref{fig:dispersion}(d). This inconsistency can be attributed to the numerical errors in the two-dimensional simulations for Fig.~\ref{fig:transmission}.

In the case of the zigzag interface, the two transmittance curves are nearly overlapped for frequencies below \SI{170}{\tera\hertz}, as shown in Fig.~\ref{fig:transmission}(d). This observation indicates that the backscattering of light at the defect is negligible, highlighting the robustness of the edge states at the zigzag interface. This outcome aligns with expectations since valley-dependent transport should exhibit resilience against valley-conserving perturbations. In contrast, the transmittance spectra of the shifted interface display a significant response to the presence of a defect. Specifically, for the shifted interface, the transmittance curve in the presence of an L3 defect is considerably lower compared to the case without the defect, as shown in Fig.~\ref{fig:transmission}(e). This discrepancy suggests that substantial backscattering has occurred at the defect, implying that the edge states at the shifted interface are less robust against this lattice defect. By comparing the transmittance spectra of the zigzag and shifted interfaces, we can conclude that the deformations in the VPC interface lead to a reduction in the robustness of the edge state against valley-conserving defects. This reduction indicates that the deformation disrupts the valley-dependent transport of the edge states.

Our work reveals that the existence of valley-dependent gapless edge states depends on the interface geometry of VPCs. By deforming a VPC interface while preserving the conservation of valleys, we observe a transition of edge states from gapless to gapped, which strongly suggests the breaking of topological protection. Furthermore, we observe the occurrence of slow light within the BZ, which is distinct from the typical resonant zero-group-velocity modes found at the BZ edge. We also provide strong evidence for the disruption of valley-dependent transport of these edge states. Our results indicate that the valley-dependent gapless edge states are not protected by valley-dependent topology alone. Instead, interface geometry is a critical factor in engineering edge states in VPCs.

\begin{acknowledgments}
The authors acknowledge fruitful discussions with Thomas Bauer, Rene Barczyk and Ewold Verhagen.
\end{acknowledgments}

\bibliography{QVHE-interface-dependence}

\end{document}